# Probing the longitudinal momentum spread of the electron wave packet at the tunnel exit

A. N. Pfeiffer[1], C. Cirelli[1*], A. S. Landsman[1#], M. Smolarski[1], D. Dimitrovski[2†], L. B. Madsen[2], U. Keller[1]

[1]Physics Department, ETH Zurich, 8093 Zurich, Switzerland
[2]Lundbeck Foundation Theoretical Center for Quantum System Research, Department of Physics and Astronomy, Aarhus University, 8000 Aarhus C, Denmark

\* cirelli@phys.ethz.ch
\# alexandra.landsman@phys.ethz.ch
† darkod@phys.au.dk

**Abstract**

**We present an ellipticity resolved study of momentum distributions arising from strong-field ionization of Helium at constant intensity. The influence of the ion potential on the departing electron is considered within a semi-classical model consisting of an initial tunneling step and subsequent classical propagation. We find that the momentum distribution can be explained by the presence of a longitudinal momentum spread of the electron at the exit from the tunnel. Our combined experimental and theoretical study provides an estimate of this momentum spread.**

In strong-field physics and attoscience, it is often assumed that tunnel ionization is the first step that induces the subsequent dynamics [1]. Also in the case when attosecond pulses are used to promote the electron in the continuum (single photon ionization), the understanding of the electron-parent ion interaction in the presence of a femtosecond laser pulse is of crucial importance to draw conclusions on various types of time delays in ultrafast experiments [2-5]. Providing a physical insight into this interaction, semiclassical models are indispensable for guiding ultrafast experiments and devising new schemes for obtaining time resolution. In the semiclassical model of strong-field ionization, the electron first escapes from the atom by tunneling, and then in a second step it follows a classical trajectory, influenced by the parent ion potential and the femtosecond laser pulse. The simplest approach is to neglect the influence of the ion potential on the electron trajectory, the so-called Simpleman's model [6-9],[10]. There are, however, several effects that cannot be explained within the Simpleman's model. Examples of such effects are Coulomb focusing [11, 12], low energy structures at mid-infrared laser wavelengths [13, 14], and Coulomb asymmetry in above-threshold ionization [15],[16]. These effects are mostly subtle, but as the precision of experiments in attoscience increases [2, 4], so does the demand for precision in the modelling and hence more details need to be accounted for by theory [3]. Another reason why improvement of the semiclassical models is important are upcoming applications, such as for example time-resolved holography with photoelectrons [17], that make use of the influence of the ion potential on the electron trajectory.

Very recently, a study [3] performed with attosecond angular streaking [18] allowed for the determination of the natural coordinates of the tunneling current flow (tunneling geometry), and showed the importance of accurately accounting for the effective potential [19] in semiclassical models. The initial conditions for the propagation of the classical trajectory in turn depend on this effective potential and the way how the three-dimensional problem of an atom in a static field has been separated into a one-dimensional tunneling problem, but also on the momentum space distribution of the electron at the tunnel exit. The momentum spread of the electronic wavepacket in the direction transverse to the field has been studied some time ago [20, 21] and also more recently [22, 23]. Here we study the longitudinal momentum spread of the electronic wavepacket at the tunnel exit point; a quantity that has raised substantial interest and controversy [24, 25].

Specifically, we present ion momentum distributions recorded over a continuous scan of the ellipticity. By comparing the experimental momentum distributions with the classical trajectory Monte Carlo (CTMC) simulations based on the **T**unnel **I**onization in **P**arabolic coordinates with **I**nduced dipole and **S**tark shift (TIPIS) model [3], it is found that the momentum distributions strongly depend on the longitudinal electron momentum at the exit of the tunnel. Using the semiclassical TIPIS model and CTMC simulations, we provide an estimation of the longitudinal momentum spread of the electron at the tunnel exit. In this sense, our experiment probes the longitudinal momentum spread of the electron at the tunnel exit point.

The experimental setup is explained elsewhere [3] therefore the description here is kept brief. A COLd Target Recoil Ion Momentum Spectroscopy (COLTRIMS) [26] setup measures the ion momentum of helium ions, which is just the negative of the momentum of the electron measured in coincidence due to momentum conservation. Short laser pulses with a pulse duration of 33 fs (FWHM) at a central wavelength of 788 nm are produced by a Titanium:Sapphire based laser system. The electric field **F** of the pulses is (atomic units are used throughout the paper)

$$\mathbf{F}(t) = \sqrt{I} f(t) \left[ \frac{1}{\sqrt{\varepsilon^2 + 1}} \cos(\omega t + \varphi_{CEO}) \hat{\mathbf{e}}_x + \frac{\varepsilon}{\sqrt{\varepsilon^2 + 1}} \sin(\omega t + \varphi_{CEO}) \hat{\mathbf{e}}_y \right], \quad (1)$$

where $f(t)$ is the pulse envelope, $I$ is the peak intensity, $\omega$ is the laser frequency, $\varepsilon$ designates the ellipticity, the x-axis is the major axis of the polarization ellipse and the y-axis is the minor axis of the polarization ellipse. The Keldysh parameter [27] is defined by

$$\gamma = \frac{\omega\sqrt{2I_p}}{F_{max}} = \frac{\omega\sqrt{2I_p(1+\varepsilon^2)}}{\sqrt{I}}, \quad (2)$$

where $F_{max} = \sqrt{I}/\sqrt{1+\varepsilon^2}$ is the maximal electric field along the major axis of ellipticity, $I_p$ is the ionization potential (24.59 eV for Helium). In the experiment we vary the ellipticity at a constant intensity $I$, which is estimated to be 0.8 PW/cm$^2$. By varying the ellipticity at a constant intensity, $F_{max}$ changes and the Keldysh parameter

$\gamma$ varies from $\gamma = 0.51$ for $\varepsilon = 0$ to $\gamma = 0.73$ for $\varepsilon = 1$. The carrier-envelope-offset (CEO) phase $\varphi_{CEO}$ [28] was not stabilized. We note that in our experimental setup perfect circularly polarized pulse cannot be produced. Close-to-circularly polarized pulses are limited to about $\varepsilon = \pm 0.93$, due to the specifications of the quarter-wave plate used and the actual laser parameter.

Figure 1 shows momentum distributions for various values of the ellipticity obtained with anti-clockwise turning fields (designated by $\varepsilon > 0$). Clockwise turning fields were also employed (designated by $\varepsilon < 0$, see Fig. 2) and used to evaluate the reliability of the data: fields with different helicity streak the electron by equal amounts but in opposite directions. Considering both helicities reduces systematic errors [3].

The ion momentum distributions emerging after ionization by the short intense laser pulse differ qualitatively for linear ($\varepsilon = 0$) and circular polarization ($|\varepsilon| = 1$). For $\varepsilon = 0$ the distribution in the polarization plane is close to Gaussian with a maximum at zero both along and perpendicular to the field direction; in the case of close-to-circular polarization a torus forms around the center (Fig. 1).

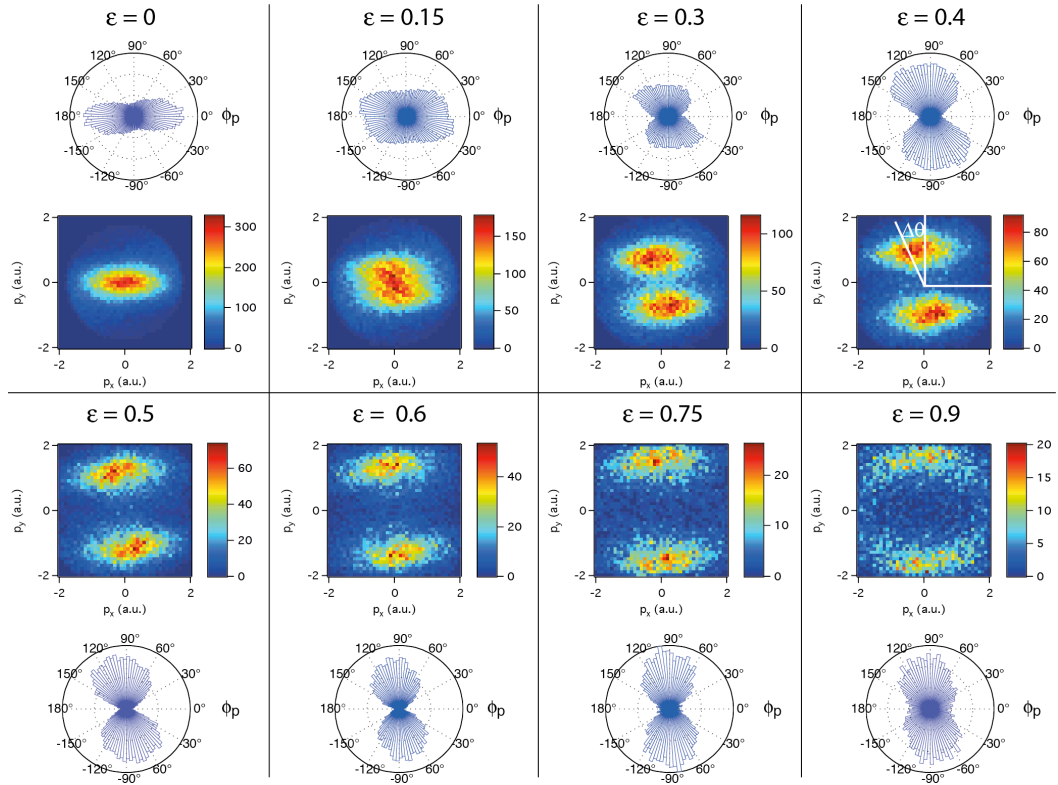

**Fig. 1. Ion momentum distributions and angular distributions in the polarization plane for different ellipticities.** The major polarization axis is the x-axis, the minor polarization axis is the y-axis, and $\phi_p = \arctan(p_y / p_x)$ is the angle of the ion momentum **p** = ($p_x$, $p_y$). For each panel, the presented data (momentum or angular distribution) is integrated over an interval of 0.05 around the ellipticity value indicated in the headline. In the panel for $\varepsilon = 0.4$ the offset angle $\Delta\theta$ is defined (see text).

An angular shift of the momentum distribution in the polarization plane in the direction of rotation of the field vector is readily seen in Fig. 1. This angular shift Δθ (defined in Fig. 1), observed in the present experiment and calculated in the tunneling regime [3], is very similar to the Coulomb asymmetry [15, 16] of individual above-threshold ionization (ATI) peaks found in the multi-photon regime. Even though in different dynamical regimes, both effects are manifestations of the influence of the atomic potential. It was shown [3] that the angular shift is very sensitive to the exact form of the parent ion potential of the target atom (see the difference for the angular offset for Helium and Argon in Ref. [3]), and on the intensity of the laser pulse. Here, another feature can be seen in Fig. 1 and that is the qualitative change in the angular distribution as a function of the ellipticity of the pulse. While the angular distribution has two distinct peaks for linear polarization and close to circular polarization, it has 4 peaks in the region around $\varepsilon = \pm 0.3$, as also predicted recently by calculations based on numerical solution of the time-dependent Schrödinger equation (TDSE) [29]. The 4 peaks are even more pronounced in Fig. 2a, where the angular distribution is shown as a spectrum resolved for the ellipticity.

For $|\varepsilon|$ larger than 0.3, and not too close to circular polarization, the angular distribution exhibits two peaks, that is, in the momentum distribution there are two lobes. The center of these lobes is shifted with respect to the minor polarization axis by the offset angle Δθ, defined in Fig. 1. This angular shift is absent in the analysis based on the Simpleman's model [10] or using the strong-field approximation [27]. However, it can be accounted for by solving the TDSE and by semiclassical calculation [3].

In our semiclassical model we perform CTMC simulations generating $10^6$ events where the time of the emission and the CEP are randomly distributed and weighted by the ionization probability. The initial conditions are the initial electron position $\mathbf{r}_0$ and initial electron momentum $\mathbf{p}_0$. Thereafter, the electron trajectory $\mathbf{r}(t)$ is obtained from the classical equations of motion

$$\frac{d^2\mathbf{r}(t)}{dt^2} = -\mathbf{F}(t) - grad\left(V(\mathbf{r},t)\right), \tag{3}$$

where $V$ is the effective atomic potential [19], i.e., $V(\mathbf{r},t) = -1/r - \alpha \mathbf{F} \cdot \mathbf{r}/r^3$ (for singly-charged parent ion) and $\alpha$ is the ionic polarizability. For Helium and at the intensity employed in this experiment, the second term in the effective potential is negligible. In all simulations we have assumed a Gaussian envelope $f(t)$ with a FWHM of 33 fs, matching the parameters of the pulse used in the experiment. The initial electron position is the point at the tunnel exit, obtained using the TIPIS model [3] at the instantaneous value of the electric field at the time of ionization. The initial momentum distribution of the electronic wavepacket will be discussed below.

To obtain the angular shift Δθ for ellipticities larger than $|\varepsilon|=0.3$, it suffices to perform a semiclassical calculation taking, in first approximation, the initial momentum equal to zero. In fact it suffices to consider only the trajectory corresponding to ionization at the maximum of the field. This can be done since there is a one-to-one correspondence between the instant of maximum ionization and the maximum in the final momentum distributions, as shown in the supplementary information to Ref. [3].

Below an ellipticity of 0.3, the angular distribution in the polarization plane changes from exhibiting two peaks into having four peaks as shown in Fig. 2a, see also Refs. [29, 30]. Then a single-trajectory semiclassical model breaks down in the range of small ellipticity and neither a model including the electron trajectories for all emission times with no initial electron momentum captures the emergence of four peaks. Since the initial condition for the position of the electron is fixed by the tunnel exit point, in the semiclassical model we can try to change the initial momentum of the electron at birth, thereby influencing the initial conditions for Eq. (3) and ultimately the final result of the simulations. We note neither the formation of four peaks in the angular distribution or the width of the actual angular distribution can be explained by the recombination of the electron wavepacket. We have checked within our semiclassical model that the probability for the recombination to happen is less than 1% of the ionization probability [31], thus this effect cannot explain our results.

As stated above, in semiclassical models the initial electron momentum is assumed to be zero in a first approximation, but our results suggest that a more detailed description of the electronic wavepacket at the tunnel exit is necessary. Indeed it is possible to model the initial momentum distribution of the electron wavepacket at the tunnel exit point, providing different initial conditions for the semiclassical simulation. The initial momentum distribution is usually discussed in terms of its components transversal and longitudinal to the polarization of the electric field. In tunneling theory and not for too strong fields [23], the momentum distribution transverse to the field direction is Gaussian centered at zero and with a standard deviation $\sigma_\perp = \sqrt{\frac{\omega}{2\gamma}}$ [20, 21]. This distribution is widely accepted and commonly used in simulations [14, 32, 33]. An experimental measurement of this distribution using circular light yielded reasonable agreement of data and theory for the wavelength used here [22]. For our parameters $\sigma_\perp$ adapts values ranging from 0.24 a.u. at $\varepsilon = 0$ to 0.20 a.u. at $\varepsilon = 1$ (dashed line in Fig. 3).

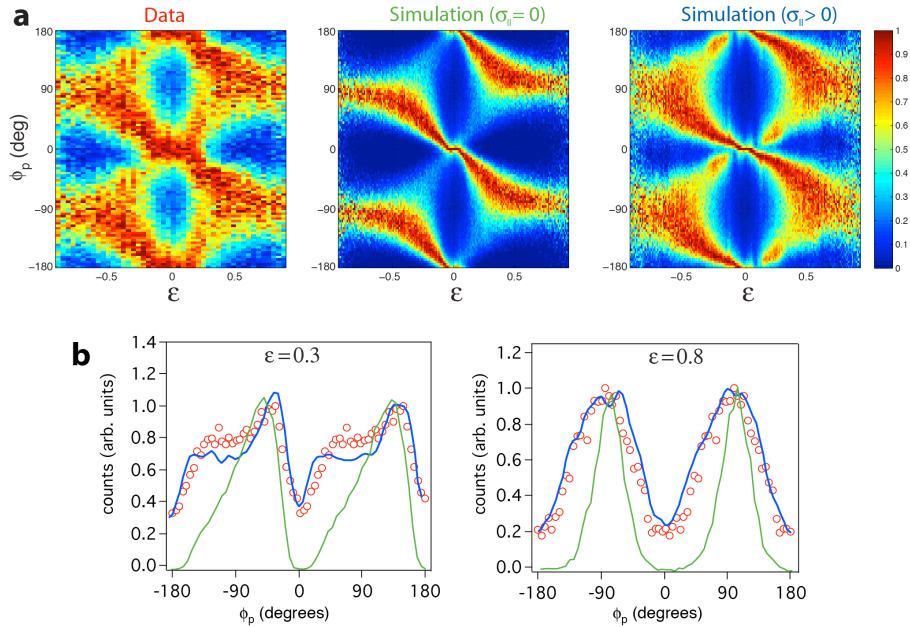

**Fig. 2. Ion angular distributions.**

a) Ellipticity-resolved angular distributions: the experiment (denoted by 'Data'), the simulation assuming initial $\sigma_\parallel = 0$, and the simulation assuming an ellipticity dependent nonzero initial $\sigma_\parallel$ longitudinal spread that yields the smallest square-error as compared to the experiment (whose values are given in Fig. 3), see text. The counts of the images are normalized for each fixed value of ellipticity in order to increase visibility,
b) Angular distributions (line cuts through a) for two different ellipticity values. The simulation with initial $\sigma_\parallel = 0$ (green curve) does not yield four peaks as in the experiment (red markers). The blue curve shows the simulation that yields the least square-error, see text and Fig. 3.

The initial longitudinal momentum distribution at the exit of the tunnel is conceptually more difficult to obtain. The difficulty stems from the fact that the laser field interacts with the ion and the electron during the tunnel ionization process [24]. A formula for the standard deviation of the momentum distribution has only been derived for the final momentum distribution after propagation in the laser field: $\sigma_\parallel^{final} = \sqrt{\dfrac{3\omega}{2\gamma^3(1-\varepsilon^2)}}$ [30]. The latter is a generalization for any ellipticity value of the standard formula for final momentum distribution in linearly polarized field, derived by the tunneling theories and introduced in Ref. [20]. Contrary to the final longitudinal spread, the initial longitudinal spread $\sigma_\parallel$ is often assumed to be zero, see for example [14]. Performing semi-classical calculations with this assumption does not reproduce the experimental data (Fig. 2).

We therefore assume that the initial momentum distribution longitudinal to the polarization is Gaussian with a standard deviation $\sigma_\parallel$ and try different values for $\sigma_\parallel$ in the semi-classical simulation. For all ellipticities we find a value for $\sigma_\parallel$ greater than zero that yields a considerably better agreement between data and simulation compared to the assumption of $\sigma_\parallel = 0$ (Fig. 2). Furthermore, we have checked that the final momentum distributions in the semiclassical model start to resemble the experiment as the longitudinal momentum spread is increased. For example, it is insufficient to take the longitudinal momentum distribution to be equal to the transverse momentum distribution, as in Ref. [12] because the four peaks cannot be obtained for $|\varepsilon| < 0.3$. Increasing the initial longitudinal momentum spread to, $\sigma_\parallel = \sqrt{3\omega/2\gamma^3}$ [20] one can obtain qualitative agreement with the experiment. However, the value of the "optimal" initial $\sigma_\parallel$ that yields the least square error as compared to the experiment is larger than the above analytic expression. This "optimal" value has been obtained as follows. For any fixed range of ellipticity, CTMC simulations with an initial longitudinal momentum spread ranging from 0 to 1.5 a.u. in steps of 0.05 a.u. are performed and the one that yields the smallest sum of squared differences with respect to the experimental data is selected.

In Fig 2b, angular distributions for selected ellipticities are given. For $\varepsilon = 0.8$, the distribution exhibits two peaks and the positions of the peaks are well reproduced by both models, even though the simulation with $\sigma_\parallel = 0$ greatly underestimates the peaks' angular width. For $\varepsilon = 0.3$, only the simulation with nonzero initial longitudinal spread captures the formation of 4 peaks in the angular distribution.

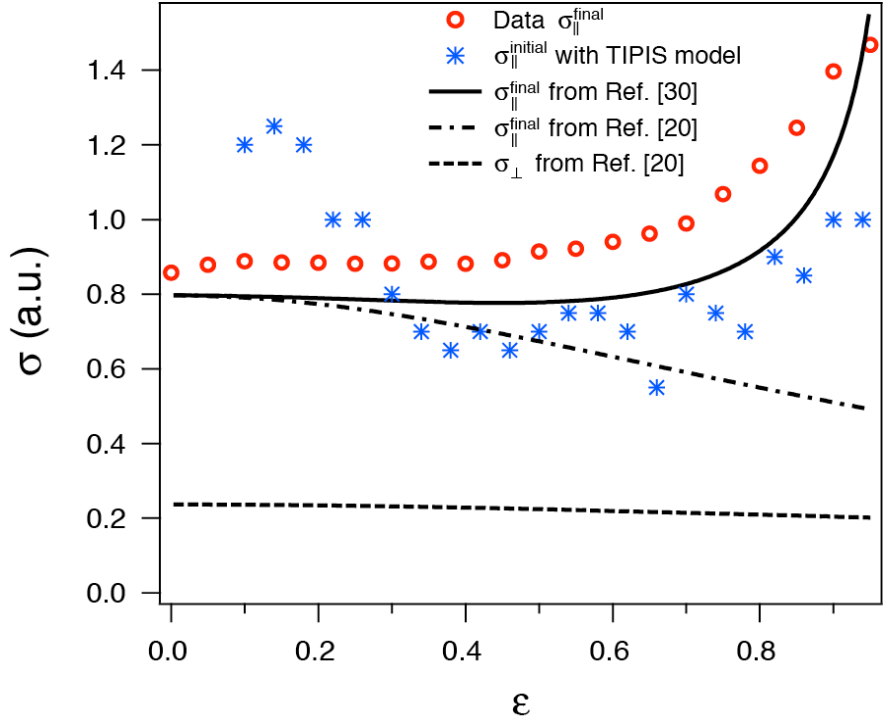

**Fig. 3**: The blue stars display the value for initial σ∥, chosen in semiclassical calculations, that yields the least square-error with respect to the experimental angular distributions from Fig. 2b). The red circles are the experimental data for the final momentum spread in the x-direction, showing a good agreement with the final longitudinal spread $\sigma_\parallel^{final} = \sqrt{\dfrac{3\omega}{2\gamma^3(1-\varepsilon^2)}}$ as a function of ellipticity predicted by Ref. [30], depicted as a solid black line. For comparison, the values for the final momentum spread in perpendicular $\sigma_\perp = \sqrt{\dfrac{\omega}{2\gamma}}$ (black dashed line) and parallel direction $\sigma_\parallel = \sqrt{3\omega/2\gamma^3}$ (black dot-dashed line) introduced in Ref. [20] are also shown.

In Fig. 3 the estimates for the initial momentum spread obtained by the semiclassical simulation are displayed together with the experimental values for the final momentum spread. The latter are deduced by a Gaussian fit of the 2D momentum distributions of Fig. 1, projected onto the x-axis (major polarization axis). The value of initial σ∥ that returns the least-square error as compared to the experiment is chosen and displayed in Fig. 3.

It is interesting to note that for |ε| < 0.3, where the formation of the asymmetric 4 peaks is the dominant feature in the momentum distribution, the final momentum spread is smaller than the initial spread at the tunnel exit. A possible reason for this is the presence of multiple returns of the liberated electron to the proximity of the ion core. This is similar to the effect of Coulomb focusing on the lateral momentum distribution as reported in Ref. [12]. For values of ellipticity larger than 0.3 the trend is reversed: the optimal momentum spread at the exit of the tunnel is smaller than the one measured at the detector, the latter being in fairly good agreement (within 10% error) with theoretical predictions.

We have presented measurements of the angular distribution of the ion momenta with a high resolution in ellipticity. Within the semiclassical model, we show that the initial spread of the electron wavepacket in the longitudinal direction has a significant impact on the angular distribution of the ion momenta. The good agreement between the experimental data and the semiclassical model provides an evidence of the existence and enables an access to the momentum spread of the electronic wave packet in the longitudinal direction at the tunnel exit point as a function of ellipticity.

Comparing the initial and the final longitudinal momentum spread, we can observe a crossover between the two quantities at ellipticity about 0.3. This is the same critical value where the qualitative change in the momentum distributions happens: the one-to-one correspondence between the timing of maximum probability of emission and the peak of momentum distributions is lost, even when the classical trajectory corresponding to the maximum emission does not rescatter. This raises deeper questions on the validity of classical trajectory approaches near and in the rescattering regime, and has implications not only for single ionization, but also for the modeling of the experiments on sequential double ionization [34-36].

## Acknowledgements

This work was supported by the NCCR Quantum Photonics (NCCR QP) and NCCR Molecular Ultrafast Science and Technology (NCCR MUST), research instruments of the Swiss National Science Foundation (SNSF), and by ETH Research Grant ETH-03 09-2 and an SNSF equipment grant and by the Danish Research Council (Grant No. 10-085430).